\documentstyle[pra,aps,multicol,psfig]{revtex}
\begin{document}
\tighten
\draft
\title{Monte-Carlo calculation of longitudinal and transverse resistivities
in a model Type-II superconductor}
\author{T.J.~Hagenaars$^*$\\
{\it Institut f\"ur Theoretische Physik, Universit\"at W\"urzburg,\\
Am Hubland, 97074 W\"urzburg, Germany}}
\author{E.H.~Brandt\\ 
{\it Max-Planck-Institut f\"ur Metallforschung, Institut f\"ur Physik,\\
Postfach 800665, 70506 Stuttgart, Germany\\}}
\date{\today}
\maketitle
\begin{abstract}
We study the effect of a transport current  on the 
vortex-line lattice in isotropic
type-II superconductors
in the presence of strong thermal fluctuations 
by means of 'driven-diffusion' 
Monte Carlo simulations of a discretized London theory
with finite magnetic penetration depth.
We calculate the current-voltage (I-V) characteristics
for various temperatures, for transverse as well as
longitudinal currents $I$.
From these characteristics, we estimate the linear
resistivities $R_{xx}=R_{yy}$ and $R_{zz}$ 
and compare these with equilibrium results
for the vortex-lattice structure factor and the helicity moduli.
From this comparison a consistent picture arises, in which
the melting of the flux-line lattice occurs in two stages
for the system size considered.
In the first stage of the melting, at a temperature $T_m$,
the structure factor drops to
zero and $R_{xx}$ becomes finite.
For a higher temperature $T_z$, the second stage takes place, 
in which the longitudinal superconducting coherence is lost, and
$R_{zz}$ becomes finite as well.
We compare our results with related recent numerical work and
experiments on cuprate superconductors.
\end{abstract}
\begin{multicols}{2}
\section{Introduction}
The statistical mechanics of vortex lines in type-II superconductors has been
the subject of intense study after the discovery of the high-temperature
superconductors.  Due to the large thermal fluctuations and pronounced
anisotropies of these materials, thermal wandering of vortex lines leads
to a rich phase diagram \cite{Blatter,Brandt}.
Numerous experiments \cite{manyexpts,YBmag,BiSmag,Schilling,Doyle}
and computer simulations \cite{manysims,Hetzel,CaCaval,Li,taochen,chen97XY,chen97,Dominguez,Nguyen,Nordborg,RyuStroud,Nguyen2,Koshelev,Hu}
were interpreted as evidence for the scenario of a vortex-lattice melting
transition into a vortex-liquid phase. Most convincing evidence for the
existence of a first-order phase transition separating the vortex 
lattice from a liquid
phase comes from a recent calorimetric measurement of the specific heat 
in YBCO \cite{Schilling} and from magnetization measurements
in YBCO \cite{YBmag} as well as BiSCCO \cite{BiSmag}. An explanation
of the magnitude and temperature dependence of the observed
characteristic entropy and magnetization jumps has been given 
in a recent work by Dodgson {\em et al.} \cite{MaDo}.

One interesting aspect of
the melting transition is the question of whether 
it coincides with a complete loss of  c-axis correlation
(i.e. decoupling of the vortex lines into independent
'pancakes' in the Cu-O layers). A recent transport measurement 
on untwinned YBCO by
Righi {\em et al.} \cite{Righi} indicates that just above the melting
temperature, the vortices are still correlated over a few microns,
and become fully decoupled only at a distinctly
higher temperature. 

On the theoretical side,  the properties of the 
vortex liquid phase have been intensively studied.
Some time ago Li and Teitel \cite{Li} numerically studied
the (uniformly frustrated) 3D XY model on a cubic lattice, and found 
a melting transition into
a liquid with longitudinal superconducting coherence.
The longitudinal superconductivity, signalling a c-axis
vortex correlation over the full system thickness,
was found to be lost at a
distinct temperature above the melting temperature.
In later work including
screening effects using the lattice London Model, a similar
two-stage melting transition was found by Chen and Teitel 
\cite{taochen,chen97}.
Recently, a two-stage transition was also found by
Ryu and Stroud \cite{RyuStroud} using a 3D XY model on a
stacked triangular lattice, for a frustration lower than 
studied in earlier work by Hetzel {\em et al.} \cite{Hetzel}. 
When the intermediate liquid  with longitudinal coherence
would persist into the
thermodynamic limit (very large system thickness), it would be
the realization of the so-called line-liquid phase proposed
in Ref. \cite{Feigelman}.

In this paper we study the dynamical properties of the lattice
London model and compare them with the behavior of equilibrium
quantities as studied in Refs. \cite{taochen,chen97,Nguyen,longsup}.

\section{Lattice London Model and Monte Carlo method}

The Hamiltonian of the isotropic lattice London model, at constant
induction $B$, reads:
\begin{equation}
{\cal H}=4\pi^2 J\sum_{i,j,\mu}q_\mu(\bbox{R}_i)q_\mu(\bbox{R}_j)g(\bbox{R}_i-\bbox{R}_j)
\label{Ham}
\end{equation}
where $J=\Phi_0^2 d/(32\pi^3\lambda^2)$ (with $\lambda$ the magnetic penetration depth) and $g(\bbox{R})$ 
is the  London  interaction with Fourier components
\begin{equation}
g(\bbox{k})=\frac{1}{\kappa^2+ (d/\lambda)^2} 
\label{Fcom}
\end{equation}
and $\kappa^2=\sum_\mu \kappa_\mu^2$ with 
$\kappa_\mu^2 =2-2\cos k_\mu$ ($k_\mu =2\pi n_\mu/L_\mu,$ 
$ n_\mu= 0,1,...,L_\mu -1$).
Here we assumed an $L_x\times L_y \times L_z$ lattice with  
periodic boundary conditions.
At every dual lattice site
$\bbox{R}$  of our square lattice, the integer variable $q_\mu(\bbox{R})$
denotes the vorticity or number of flux-line unit elements
in the direction $\mu =x,y,z$.  
The $q_\mu (\bbox{R}_i)$ are subject to the continuity constraint
$\sum_{\bbox{e}_\mu}\left[ q_\mu (\bbox{R}_i)-q_\mu (\bbox{R}_i-\bbox{e}_\mu)\right] =0.$
Here $\bbox{R}_i-\bbox{e}_\mu$ runs over
nearest neighbor sites of $\bbox{R}_i$.
$\lambda$ is the magnetic penetration depth
and $d$ the lattice constant.
The Hamiltonian (\ref{Ham}) can be derived
from the discrete version of the London free energy \cite{CaCaval,chen97,CaPRB94}.

Monte Carlo sampling of the phase space for the variables
$q_\mu(\bbox{R})$  at constant  $B$
is performed as follows.
In the simulations, the initial configuration is prepared to contain the number of vortex
lines corresponding to the value chosen for $B$.
A Monte Carlo 
update step consists of adding at a given site a closed
$d\times d$ square loop of unit vorticity with an orientation chosen
randomly from the six possible ones.
This scheme preserves the magnetic induction 
$B$ with components
$B_\mu= \frac{\Phi_0}{d^2 {\cal V}}\sum_j \langle q_\mu (\bbox{R}_j)\rangle .$
(${\cal V}=L_xL_yL_z$).
The standard Metropolis algorithm is employed to accept or reject
the new configuration.  For equilibrium calculations, one uses
(\ref{Ham}) to calculate the energy change $\Delta E$ between the old and
the new configuration. To simulate the presence of a uniform transport 
current $\bbox{j}$, we introduce an additional bias in the acceptance rates by
subtracting or adding
$(\Delta E)_j= \Phi_0 \mid \bbox{j}\mid  d^2/c$ to the energy change
$\Delta E$ when  an  elementary loop is added whose normal vector is pointing
in the same direction as $\bbox{j}$ or $\bbox{-j}$ respectively.
We measure the  magnitude of the current in terms of the
dimensionless quantity $\bbox{\alpha}\equiv \bbox{j}\Phi_0d^2/(Jc)$.
The dimensionless voltage $V_\mu$ is 
obtained by measuring the  rate  at which
vortex jumps  in the $\mu$ direction are occurring.
This 'driven-diffusion' method was used before in Refs.\cite{Wallinetc}.

\section{Results }
Our calculations  are carried out on a $15\times 15\times 15$
cubic lattice, with a magnetic field running in the
$z$ direction. We choose the filling fraction
$f=1/15$ and $\lambda=5d$, as in
Refs. \cite{taochen,chen97}.  Runs were taken consisting of
16,384 (high currents) up to 262,144 (low currents)
Monte Carlo  sweeps through the lattice, half of which
were used for equilibration.
\begin{figure}
\centerline{}
\centerline{\hspace{1.0cm}
\psfig{figure=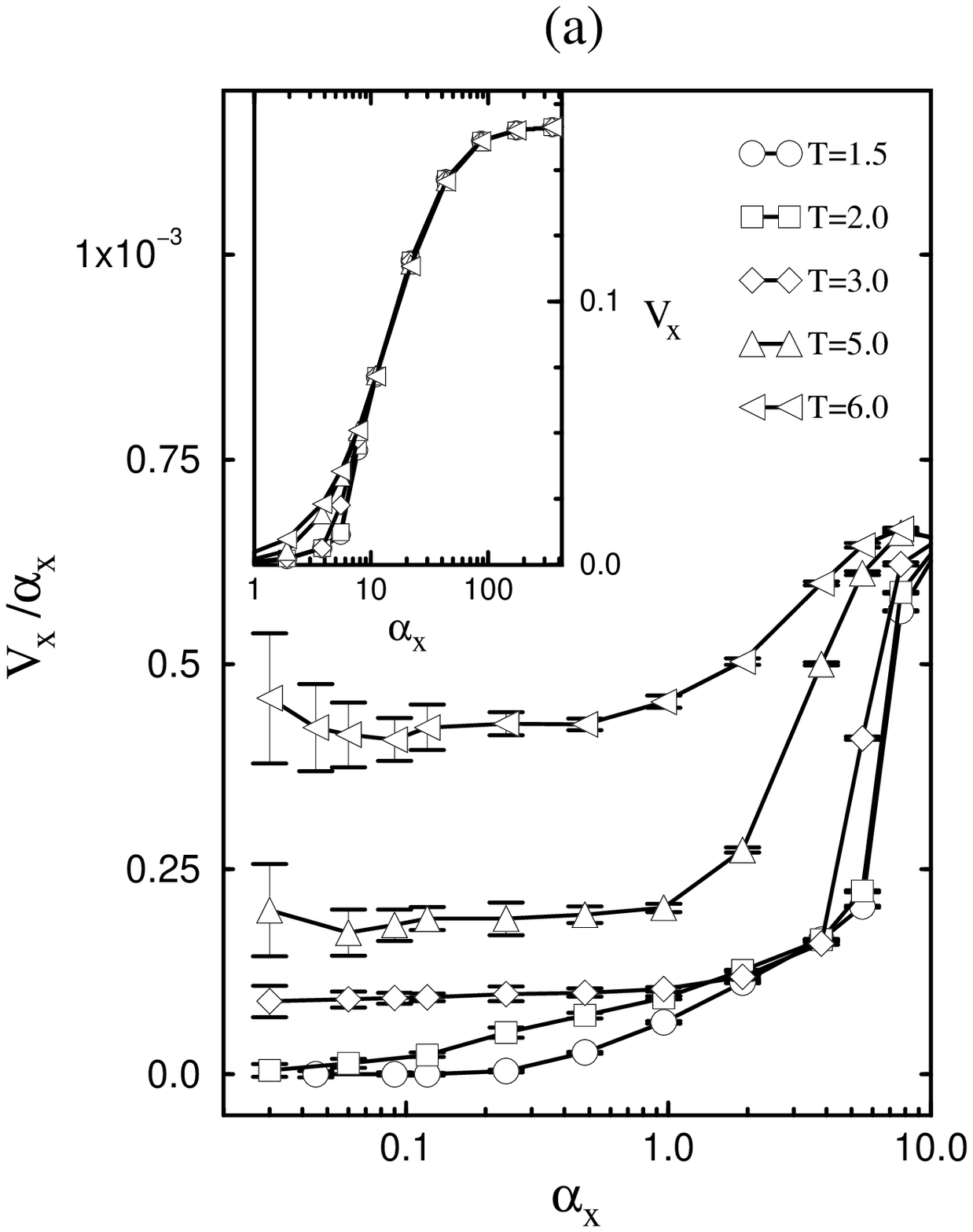,height=9.0cm,width=11.4cm}}
\centerline{}
\centerline{}
\centerline{\hspace{1.0cm}
\psfig{figure=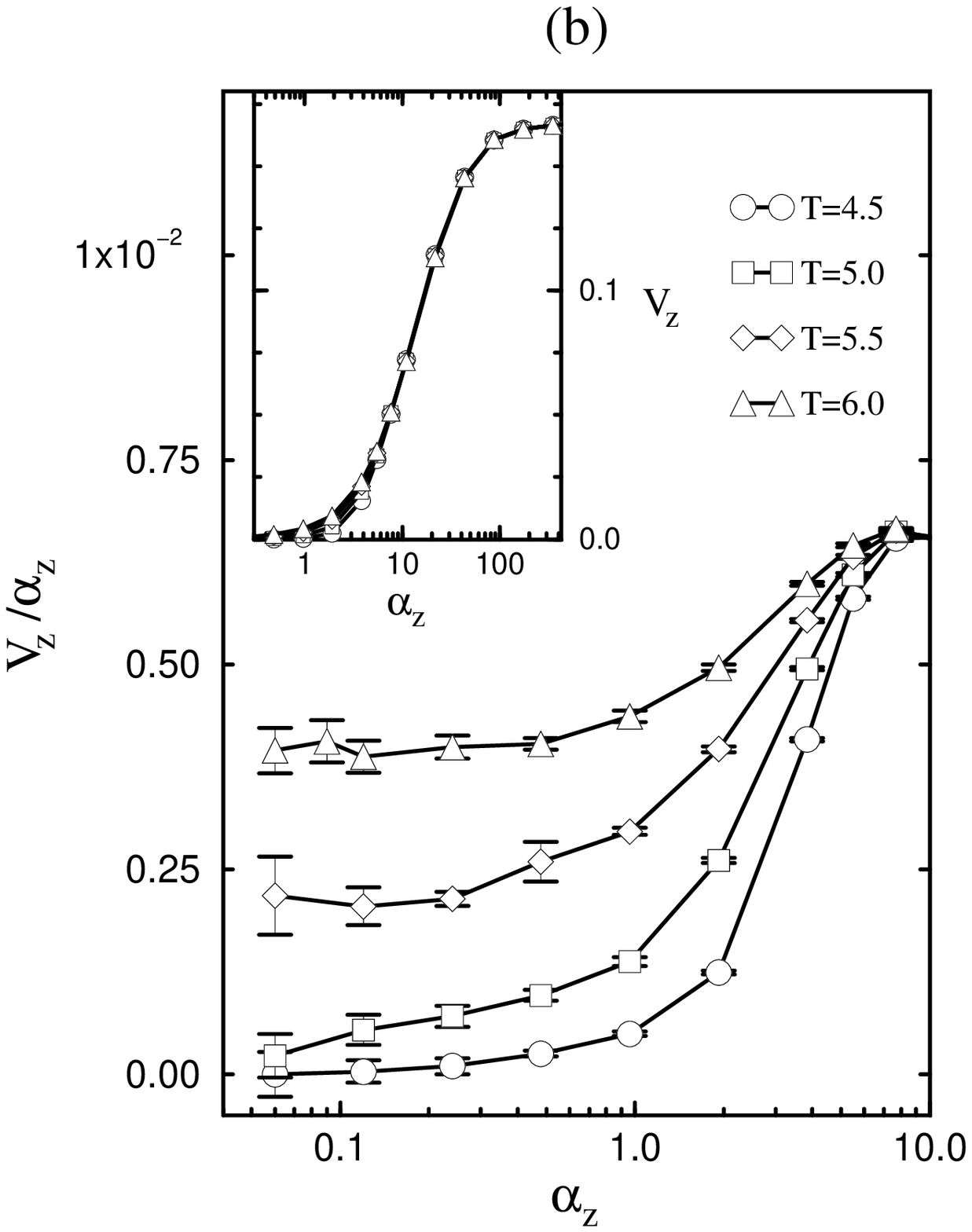,height=9.0cm,width=11.4cm}}
{FIG.~1. \small I-V characteristics, plotted as resistance versus
current. (a) current along $x$ direction.
(b) current along $z$ direction. Insets: I-V characteristics, plotted
as voltage versus current, including
the high current range.}
\label{fig1}
\end{figure}
In Fig. 1(a) and (b) we show our results for the I-V characteristics
at different temperatures, plotting the resistances
${V}_\mu /\alpha_\mu$ as a function of
current $\alpha_\mu$ for $\mu=x$ and $\mu=z$, respectively.
In Fig. 1(a) the in-plane ($\mu=x$) resistance is shown.
For the lowest T shown ($T=1.5$ in units of $J/k_B$), the resistivity drops
to zero below a finite current value (critical current),
and thus the linear resistivity
\begin{equation}
R_{xx}\equiv \lim_{\alpha_x\rightarrow 0} \frac{V_{x}}{\alpha_x}
\end{equation}
is zero.
For $T=2.0$ no clear critical current is observed, but again
we estimate $R_{xx}$ to be zero (or very small).
For the higher $T$'s studied,
we find that the
I-V characteristic is linear
(constant resistance) for sufficiently small currents.
For high currents,  the I-V characteristics become independent
of temperature (see insets in Figs. 1(a) and (b)). 
In the high-current limit the voltage saturates at the value 
$1/6$, corresponding to acceptance of all the Monte Carlo trial moves
in the direction of the Lorentz force.
\begin{figure}
\centerline{
\psfig{figure=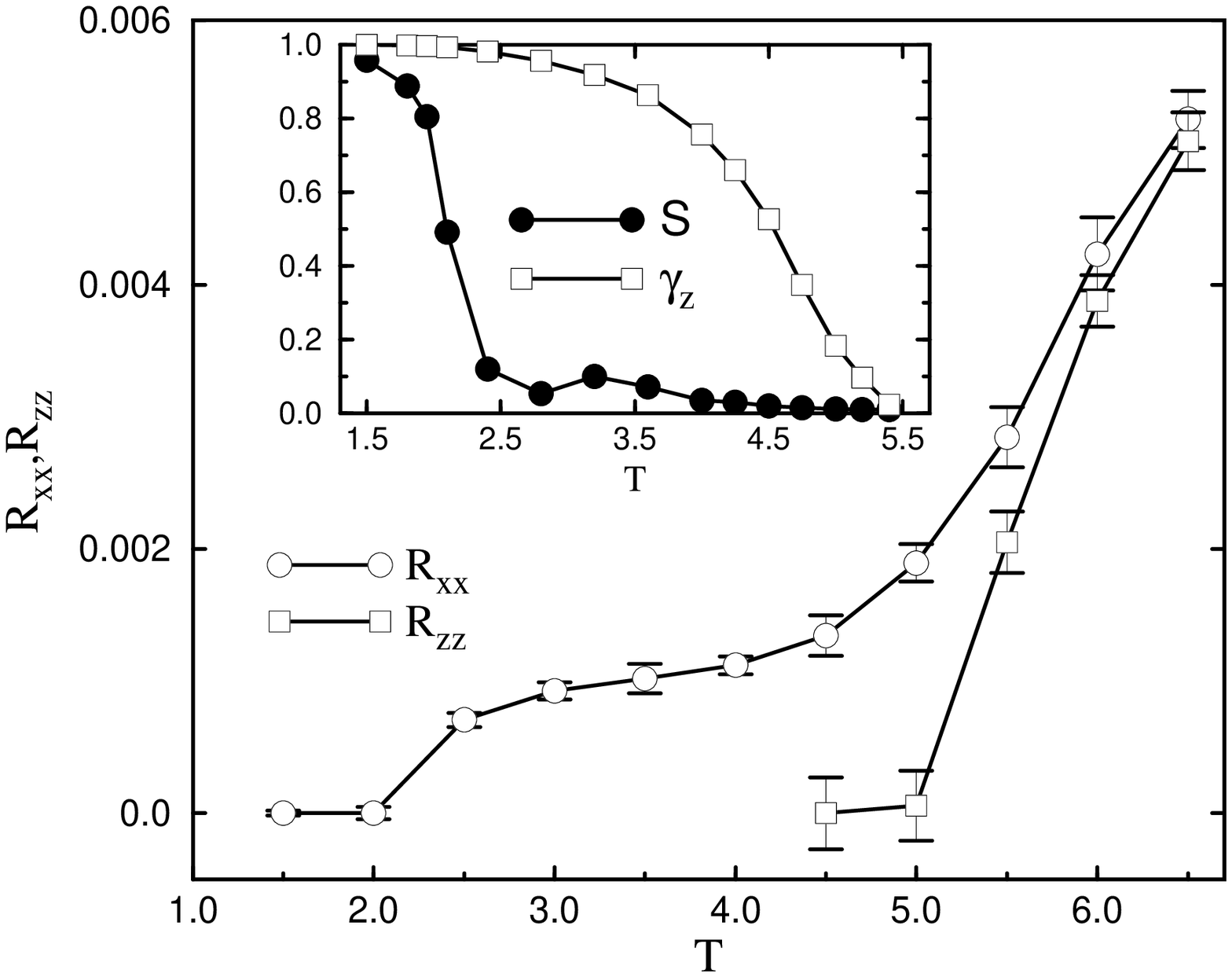,height=7.8cm,width=9.0cm}}
{FIG.~2. \small Linear resistivities vs. T. Circles: in-plane.
Squares: out-of-plane. Inset: temperature dependence of the
vortex-lattice structure factor $S\equiv S(\bbox{k}_s)$, with $k_s$ the smallest
reciprocal lattice vector of the ground state vortex 
lattice, and the
helicity modulus $\gamma_z$, calculated at $\bbox{j}=0$.}
\label{fig2}
\end{figure}
In Fig. 2 we plot the linear resistivities $R_{xx}$ and $R_{zz}$
(estimated from the I-V characteristics of which representative
ones are shown in Fig. 1)
as a function of $T$. 
We distinguish three temperature regions with different
dissipative properties. For low $T$, both $R_{xx}$ and $R_{zz}$
are zero. This is the flux-line lattice state, stabilized
against small currents by the artificial pinning effect
of the discrete mesh in our simulation.
For intermediate $T$, $R_{xx}$ becomes finite but $R_{zz}$ is
still zero. Finally, for high $T$, both transport coefficients
become nonzero.
For comparison, we include equilibrium (i.e. $\bbox{j}=0$) results
for the structure factor $S$ and  the helicity
modulus $\gamma_z$ in the inset of Fig. 2. 
These equilibrium results were 
obtained before in Refs. \cite{taochen,chen97}.
The temperature at which the in-plane linear resistance
becomes nonzero coincides with the
melting temperature $T_m$ of the flux-line lattice as found
from the decay of the equilibrium structure factor 
\cite{opmerk}.
Furthermore,  the temperature at which the out-of-plane linear
resistance $R_{zz}$ becomes nonzero coincides
with the temperature $T_z$ at which the 
$\bbox{j}=0$ helicity modulus $\gamma_z$ 
vanishes (see inset). Here $\gamma_z$ was calculated
as explained in detail in Ref. \cite{longsup}.
This is consistent with the interpretation of $\gamma_z$ as a
measure of longitudinal superconducting
coherence or longitudinal superconductivity. 
Thus our transport calculations confirm the  two-stage 
melting picture found in
equilibrium simulations of this model \cite{taochen,chen97}, in which 
there is an intermediate vortex-liquid regime  with superconductivity
along the vortex lines.
In this intermediate regime $T_m<T<T_z$ the vortex lines are
c-axis correlated over the full system thickness
$L_z=15$ simulated.  
For $T>T_z$ the longitudinal correlation
length becomes smaller than $L_z$. At some even higher $T_d$
(decoupling temperature),
the c-axis correlation should be lost completely.

\section{Discussion}

As our results are obtained for a finite system size, 
it is not clear from them whether the two-stage melting
scenario persists into the thermodynamic limit.
In fact, we believe that this will not be the case.
In Refs. \cite{taochen,chen97} it was found that the
$T_z$ decreased with increasing system thickness $L_z$. 
In Ref. \cite{Nguyen2}, the $L_z$ dependence of $T_z$
was studied for the uniformly frustrated Villain Model.
There it was also found that the intermediate temperature
region decreased in size with increasing $L_z$.
We interpret this finite-size behavior as a
manifestation of a c-axis correlation length that, in 
a thick system, is finite and large just above the melting
temperature and shrinks with increasing temperature
until a decoupling temperature $T_d$ is reached.
Thus, what these simulation results {\em do} predict
for real (i.e. thick)  systems is that the vortex lattice
melts into a liquid in which the vortex lines are
c-axis correlated over many layer distances, i.e.
far from decoupled.
Only for higher temperatures a crossover takes place
into a decoupled regime above $T_d$.
This scenario agrees well with the transport experiments 
on YBCO by
Righi {\em et al.} \cite{Righi}, that indicate that just above the melting
temperature, the vortices are still correlated over a few microns
\cite{Trawick},
i.e. thousands of
Cu-O layer distances, and become fully decoupled only at a distinctly
higher temperature. 
In contrast, measurements on BiSCCO 
by Doyle {\em et al.} \cite{Doyle}
were interpreted as evidence for a simultaneous melting and decoupling.
This is consistent with recent Monte-Carlo simulations of a 
strongly anisotropic modified 3D XY model by Koshelev \cite{Koshelev}, 
in which it was found that 
the longitudinal helicity modulus drops to zero close to the 
melting temperature already for $L_z=40$ and larger.
We note that such a behavior was also found for $L_z=40$ in
a recent Monte Carlo study of an only slightly anisotropic
3D XY model, in which the melting transition
 was investigated upon cooling \cite{Hu}.


We thank G. Carneiro, W. Hanke, S. Teitel, and M. Trawick for discussions.
We gratefully acknowledge financial 
support by the ``Bayerischer Forschungsverbund Hochtemperatur-Supraleiter (FORSUPRA)".

\end{multicols}

\begin{references}
\bibitem{opmerk}We note that, due to the discrete-mesh pinning effect,
the melting temperature $T_m$ is actually an upper bound for the
true melting temperature. Ideally,
the vortex lattice should depin from the numerical mesh below $T_m$. 
Such a behavior can be achieved by softening the artificial 
pinning effect using a more dilute vortex system (see e.g. Ref. 
\cite{Nguyen}).
\end{references}
\end{document}